\documentclass[showpacs,preprintnumbers]{revtex4}
\usepackage{amsmath}
\usepackage{graphicx}

\begin{document}

\textbf{Comment on \textquotedblleft Macroscopic Test of the Aharonov-Bohm
Effect\textquotedblright \bigskip }

Tomislav Ivezi\'{c}

\textit{Ru%
\mbox
{\it{d}\hspace{-.15em}\rule[1.25ex]{.2em}{.04ex}\hspace{-.05em}}er Bo\v{s}%
kovi\'{c} Institute, P.O.B. 180, 10002 Zagreb, Croatia}

\textit{ivezic@irb.hr\bigskip }

In [1] the absence of forces for the magnetic Aharonov-Bohm (AB) effect [2]
has been experimentally investigated by means of a time-of-flight experiment
for a macroscopic solenoid. It is looked for a time delay for electrons
passing on opposite sides of the solenoid. In the generally accepted theory
the AB effect is considered to be a purely quantum mechanical in nature. The
electron wave packets are influenced by nonzero vector potential, i.e., by
the quantum action of the magnetic flux, even when electrons pass through
the field-free regions of space, Eq. (1) and Fig. 1(a) in [1]. On the other
hand in Boyer's semiclassical theory, Ref. [16] in [1], there is a
back-action force of the solenoid on the electron, which gives rise to a
time delay, Eq. (4) in [1], and to a phase shift, Eq. (8) in [1], that
exactly matches the AB-phase shift. As shown in [1] \textquotedblleft no
time delay is observed (Fig. 3), thus signaling the absence of
forces.\textquotedblright\ But, in [3], Boyer stated: \textquotedblleft the
Aharonov-Bohm phase shift has never been observed for such a macroscopic
solenoid, .. .\textquotedblright\ In [3], it is also argued that if the
solenoid resistance is large, as in [1], then the back forces will be small
and there is no time lag, but for the microscopic solenoids it is the
opposite case.

However, as explained in Sec. 10 in [4], in the experiments from [1] and
Ref. [6] in [1], in all theoretical discussions including [2] and Boyer's
semiclassical theory, it is \emph{never} noticed that \emph{always} there is
an electric field outside stationary resistive conductor carrying constant
current. In such ohmic conductor there are quasistatic surface charges that
generate not only the electric field inside the wire driving the current,
\emph{but also a static electric field outside it}, which has nothing to do
with Boyer's force picture. There are no analytic solutions for these
surface charges and the external electric fields for the case of finite
solenoids; for an infinite solenoid see [5]. For the hystorical analysis and
for some experimental confirmation see Ref. [42] in [4]. The distribution of
the surface charges and the magnitude of the induced electric fields depend
not only on the geometry of the circuit but even of its surroundings. These
fields are well-known in electrical engineering, which means that they can
be much bigger than those in Boyer's picture. Hence, the main result from
[1] does not imply that the electrons travel in a field-free region. These
fields have to be taken into account for the explanation of the AB phase
difference even in the magnetic AB effect. A similar explanation is already
proposed in [6], Eq. (28), but their calculation is not relativistically
correct. In Secs. 7-7.2 in [4] it is shown that even if the experiments
would be made with superconducting solenoids with steady currents there
would be the external electric field. In Sec. 8 in [4] such electric fields
are predicted to exist for a \emph{stationary} permanent magnet as well.
Note that in [1] the whole treatment is with the 3D quantities. In the
recent paper [7] the covariant expression for the AB phase difference $%
\delta \alpha _{EB}$\ in terms of the Faraday 2-form $F$ is presented, $%
\delta \alpha _{EB}=(e/\hbar )\int F$,\ where $F=(-1/2)F_{\mu \nu }dx^{\mu
}\wedge dx^{\nu }$, $F_{\mu \nu }=(v_{\mu }E_{\nu }-v_{\nu }E_{\mu
})+\varepsilon _{\mu \nu \alpha \beta }v^{\alpha }B^{\beta }$, $E_{\mu }$
and $B_{\mu }$ are the components of the 4D electric and magnetic fields
respectively, $v_{\mu }$ are the components of the 4D velocity of a family
of observers who measure electric and magnetic fields, see also [4]. If the
observers are at rest in the rest frame of the solenoid $v^{\mu }=(1,0,0,0)$%
, $E_{0}=B_{0}=0$ and the electric part $\delta \alpha _{E}$ of $\delta
\alpha _{EB}$\ is $\delta \alpha _{E}=(e/\hbar )\int
v_{0}E_{i}(x)dx^{i}\wedge dx^{0}$. There, in [7], it is also argued that in
the 4D spacetime only $\delta \alpha _{E}$ is physically correct and
justified in the magnetic AB effect, because only the electric field from
the solenoid with steady current exists in the region outside the solenoid
and it can \emph{locally} influence the electron travelling through that
region. In order to clarify the situation some new experiments are required:
the measurement in a \emph{single} experiment of the AB phase shift and the
time delay, as suggested in [3], and the measurement of the mentioned
external electric fields \emph{separately} from AB-studies.\bigskip

\noindent \textbf{References\bigskip }

\noindent \lbrack 1] A. Caprez, B. Barwick and H. Batelaan, Phys. Rev. Lett.
\textbf{99}, 210401 (2007).

\noindent \lbrack 2] Y. Aharonov and D. Bohm, Phys. Rev. \textbf{115}, 485
(1959).

\noindent \lbrack 3] T. H. Boyer, Found. Phys. \textbf{38}, 498 (2008).

\noindent \lbrack 4] T. Ivezi\'{c}, J. Phys.: Conf. Ser. \textbf{437},
012014 (2013); arxiv 1204.5137.

\noindent \lbrack 5] M. A. Heald, Am. J. Phys. \textbf{52}, 522 (1984).

\noindent \lbrack 6] H. Torres S., A. K. T. Assis, Revista de la Facultad de
Ingenieria de la

Universidad de Tarapaca (Chile) \textbf{9}, 29 (2001).

\noindent \lbrack 7] T. Ivezi\'{c}, submitted to Phys. Lett. B.

\end{document}